%% file: main.tex
\documentclass[sigconf]{acmart}
\copyrightyear{2021}
\acmYear{2021}
\setcopyright{acmcopyright}\acmConference[ESEC/FSE '21]{Proceedings of the 29th ACM Joint European Software Engineering Conference and Symposium on the Foundations of Software Engineering}{August 23--27, 2021}{Athens, Greece}
\acmBooktitle{Proceedings of the 29th ACM Joint European Software Engineering Conference and Symposium on the Foundations of Software Engineering (ESEC/FSE '21), August 23--27, 2021, Athens, Greece}
\acmPrice{15.00}
\acmDOI{10.1145/3468264.3468538}
\acmISBN{978-1-4503-8562-6/21/08}

\usepackage[frozencache,cachedir=.]{minted}
\usepackage{multirow}
\usepackage{amsmath}
\usepackage{amsfonts}
\usepackage{framed}
\usepackage{bm}
\usepackage[skins]{tcolorbox}
\usepackage{tikz}
\usetikzlibrary{backgrounds}
\usepackage[framemethod=TikZ]{mdframed}
\usepackage{listings}
\usepackage{color}
\usepackage{algorithm}
\usepackage[noend]{algpseudocode}
\usetikzlibrary{shadows}
\usepackage{flushend}
\usepackage{xspace}

\usepackage{pbox}
\usepackage{balance}
\usepackage{soul}
\usepackage{wrapfig}
\usepackage{textcomp}
\usepackage{hyperref}
\usepackage{balance}
\usepackage{cleveref}
\usepackage{soul}

\setminted{style=borland,fontsize=\small}

\definecolor{dkgreen}{rgb}{0,0.6,0}
\definecolor{gray}{rgb}{0.5,0.5,0.5}
\definecolor{mauve}{rgb}{0.58,0,0.82}
\definecolor{codegreen}{rgb}{0,0.6,0}
\definecolor{codegray}{rgb}{0.5,0.5,0.5}
\definecolor{codepurple}{rgb}{0.58,0,0.82}
\definecolor{backcolour}{rgb}{0.95,0.95,0.92}

\newcommand{\etal}{et al. }

\newcommand{\fref}[1]{Figure~\ref{fig:#1}}
\newcommand{\sfref}[1]{~\ref{fig:#1}}
\newcommand{\tref}[1]{Table~\ref{tab:#1}}

\newcommand{\sect}[1]{~\cref{sect:#1}}

\newcommand{\mintJ}[1]{\mintinline{Java}{#1}}
\newcommand{\mintP}[1]{\mintinline{Python}{#1}}

\newcounter{resQues}
\newcommand{\theResQues}{\arabic{resQues}}

\newenvironment{resQues}[1][]{%
 
    \ifstrempty{#1}%
    {
    \refstepcounter{resQues}
    \mdfsetup{%
        frametitle={%
            \tikz[baseline=(current bounding box.east),outer sep=2pt]
            \node[anchor=east,rectangle,fill=white]
            {\strut \textbf{RQ~\theResQues}};}
        }%
    }{\mdfsetup{%
        frametitle={%
            \tikz[baseline=(current bounding box.east),outer sep=2pt]
            \node[anchor=east,rectangle,fill=white]
            {\strut \textbf{Research Question ~#1}};}%
        }%
    }%
    \mdfsetup{%
        innertopmargin=5pt,linecolor=black!90,%
        linewidth=1pt,topline=true,%
        frametitleaboveskip=\dimexpr-\ht\strutbox\relax%
    }
\begin{mdframed}[]\vspace{-4mm}\relax}{%
\end{mdframed}}

\newcounter{todoCounter}
\newenvironment{TODOEnv}[1][]{
    \refstepcounter{todoCounter}
    \noindent \textbf{TODO~\arabic{todoCounter}: ~#1}
}

\newcommand{\added}[1]{{#1}}

\newcommand{\approach}{COSAL\xspace}
\newcommand{\longapproach}{Code-to-Code Search Across Languages\xspace}
\newcommand{\slacc}{SLACC\xspace}
\newcommand{\dToken}{$d_{token}$\xspace}
\newcommand{\dAST}{$d_{AST}$\xspace}
\newcommand{\dIO}{$d_{IO}$\xspace}
\newcommand{\approachToken}{\approach{}$_{token}$\xspace}
\newcommand{\approachAST}{\approach{}$_{AST}$\xspace}
\newcommand{\approachStatic}{\approach{}$_{static}$\xspace}
\newcommand{\approachSLACC}{\approach{}$_{SLACC}$\xspace}
\newcommand{\weightedApproach}{KD$_{IO+AST+token}$}
\newcommand{\facoy}{FaCoY\xspace}
\newcommand{\clcdsa}{CLCDSA\xspace}

\newcommand{\astLearner}{ASTLearner\xspace}

\crefname{section}{\S}{\S\S}
\Crefname{section}{\S}{\S\S}
\newtcolorbox{slimbox}{top=0mm, bottom=0mm, left=0mm, right=0mm}

\usepackage{tabularx}
\usepackage{booktabs}
\usepackage{array}
\usepackage{multirow}

\newcolumntype{L}[1]{>{\raggedright\arraybackslash}p{#1}}

\newcolumntype{C}[1]{>{\centering\arraybackslash}p{#1}}

\newcolumntype{R}[1]{>{\raggedleft\arraybackslash}p{#1}}

\usepackage{xcolor, colortbl}
\definecolor{my-blue}{RGB}{83,87,118}
\usepackage{graphicx}
\usepackage{caption,subcaption}

\usepackage{arydshln}
\usepackage{siunitx}
\newcommand{\rowSep}{\\[0.5ex]\hdashline\noalign{\vskip 0.5ex}}
\newcommand{\cRowSep}[1]{\\[0.5ex]\cdashline{#1}\noalign{\vskip 0.5ex}}

\makeatletter
\def\adl@drawiv#1#2#3{%
        \hskip.5\tabcolsep
        \xleaders#3{#2.5\@tempdimb #1{1}#2.5\@tempdimb}%
                #2\z@ plus1fil minus1fil\relax
        \hskip.5\tabcolsep}
\newcommand{\cdashlinelr}[1]{%
  \noalign{\vskip\aboverulesep
           \global\let\@dashdrawstore\adl@draw
           \global\let\adl@draw\adl@drawiv}
  \cdashline{#1}
  \noalign{\global\let\adl@draw\@dashdrawstore
           \vskip\belowrulesep}}
\makeatother

\usepackage{ifthen}

\newcommand{\JAVA}{Java}
\newcommand{\PYTHON}{Python}

\def\BibTeX{{\rm B\kern-.05em{\sc i\kern-.025em b}\kern-.08em
    T\kern-.1667em\lower.7ex\hbox{E}\kern-.125emX}}
    
\begin{document}

\title{Cross-language code search using static and dynamic analyses}

\author{George Mathew}
\affiliation{%
  \institution{North Carolina State University}
  \country{USA}
}
\email{george2@ncsu.edu}
\author{Kathryn T. Stolee}
\affiliation{%
  \institution{North Carolina State University}
  \country{USA}
}
\email{ktstolee@ncsu.edu}

\renewcommand{\shortauthors}{Mathew \etal}

\begin{abstract}
As code search permeates most activities in software development,  code-to-code search has emerged to support using code as a query and retrieving similar code in the search results. 
 Applications include duplicate code detection for refactoring, patch identification for program repair, and language translation. 
Existing code-to-code search tools rely on static 
similarity approaches such as the comparison of tokens and abstract syntax trees (AST) to
approximate dynamic behavior, leading to low precision. 
Most tools do not support cross-language code-to-code search, and those that do, rely on machine learning models that require labeled training data. 

We present \longapproach{} (\approach{}), a cross-language technique that uses both static and dynamic analyses to identify similar code and does not require a machine learning model. 
Code snippets are ranked using non-dominated sorting based on code token similarity, structural similarity, and behavioral similarity. 
We empirically evaluate \approach{} on two datasets of 43,146 \JAVA{} and \PYTHON{} files and 55,499 \JAVA{} files and find that 1)~code search based on non-dominated ranking of static and dynamic similarity measures is more effective compared to single or weighted measures; and 2)~\approach{} has better precision and recall compared to state-of-the-art within-language and cross-language code-to-code search tools. 
We explore the potential for using \approach{} on large open-source repositories and discuss scalability to more languages and similarity metrics, 
providing a gateway for practical, multi-language code-to-code search.


\end{abstract}



\maketitle
\section{Introduction}
Code-to-code search describes the task of using a code query to search for similar code in a repository. This task is particularly challenging when the query and results belong to different languages due to syntactic and semantic differences between the languages~\cite{deissenboeck2012challenges}.
Consider the case of code migration, where it is common for applications in a specific language to be re-written to another language~\cite{mayer2017multi}. 
For example, while porting the video game \emph{Fez} from Microsoft XBox to Sony PlayStation, the developers faced their  biggest challenge in converting the original C\# code to C++ as the PlayStation did not support the C\# compiler~\cite{fezXBox}. 
Code-to-code search is also involved in identifying code clones~\cite{perez2019cross,ragkhitwetsagul2019siamese}, finding translations of code in a different language~\cite{nguyen2017exploring}, program repair~\cite{reiss2009semantics,barr2014plastic}, and supporting students in learning a new programming language~\cite{acar2016you}.
The growing prominence of large online code repositories  and the repetitive nature of source code~\cite{sadowski2015developers,lopes2017dejavu} lead to the presence of large quantities of potentially similar code across languages, providing a viable platform for code-to-code search.



We propose the first cross-language code-to-code search approach with dynamic and static similarity measures. The novelty is in the application of non-dominated sorting~\cite{deb2002fast} to code-to-code search, allowing static and dynamic information (without aggregation) to identify search results. \approach{} leverages prior art in clone detection using input-output (IO) behavior~\cite{mathew2020slacc}. As dynamic clone detection requires executable code, individually it cannot achieve the recall required for practical search applications. This is where the prior art in static analysis shines; we use token-based and AST-based measures to complement the dynamic analysis. \approach{} reaps the benefits of dynamic analysis in finding code that behaves similarly, when dynamic information is available, and the benefits of static information when dynamic information is infeasible. It provides results that balance \emph{how code looks} with \emph{how it behaves}, in the spirit of returning code that looks more natural to the user.

We evaluate \approach{} using 43,146 \JAVA{} and \PYTHON{} files from AtCoder, a programming contest dataset, and 55,499 \JAVA{} files from BigCloneBench~\cite{svajlenko2015evaluating}, a \JAVA{} based clone detection benchmark. We show that combining static and dynamic analyses yields better precision and success rate compared to code search with individual or weighted analyses. \approach{} performs better in cross-language and within-language contexts compared to the state-of-the-art code search tool \facoy{} and the industrial benchmark, GitHub. \approach{} can also detect more cross-language code clones compared to \slacc{} and \clcdsa{}, the state-of-the-art code clone detection techniques.
The contributions of this work are:

\begin{itemize}
    \item the first code-to-code search approach using non-dominated sorting over static and dynamic similarity measures,
    \item an evaluation of \approach{} with state-of-the-practice cross-language code search tools, GitHub and ElasticSearch (RQ2), 
    \item an evaluation of \approach{} with state-of-the-art single-language code search technique \facoy{} (RQ3),
    \item an evaluation of \approach{} against cross-language clone detection techniques CLCDSA and \slacc{} (RQ4), and 
    \item an open-source tool that performs cross-language code search on \JAVA{} and \PYTHON{} and can be extended to other languages~\cite{mocs}.
\end{itemize}

\section{Motivation}
\label{sect:motivation}
Effective code-to-code search requires code similarity measures that cover a variety of developer concerns. Code-to-code search should preserve code behavior, and thus IO similarity from dynamic analysis is an important consideration. Prior work has shown that identifiers impact source code comprehension, especially for beginners~\cite{binkley2013impact}, and as developers must understand the code returned by search, tokens are an important consideration. Prior work in code-to-code search that relies on ASTs have seen high precision and recall~\cite{perez2019cross,kraft2008cross} suggesting that is an important consideration as well. Individually, each measure has shortcomings. Taken together, however, we show the whole is greater than the sum of its parts. 



\input{figs_code_samples_mini}

Consider the code snippets in ~\fref{codeSamples}. Three of the functions are behaviorally identical, taking an input integer and returning an array of even integers: ~\sfref{javaLoop} is a \JAVA{} function which uses a for loop; ~\sfref{javaStream} uses the stream library from \JAVA{} v8; ~\sfref{pyComp} is a \PYTHON{} function which uses a filtered ~\mintP{list-comprehension}. ~\sfref{pyOdds} is different: it is a \PYTHON{} function that takes an integer $max$ and returns a list of odd numbers between $[0, max)$. \fref{tokensAndTrees} contains AST and token information for the code snippets, discussed extensively in \sect{longapproach}.

To identify behaviorally identical \PYTHON{} code for ~\fref{javaLoop}, a code-to-code search engine should support both the programming languages \JAVA{} and \PYTHON{}. In this case, using a purely token-based approach for cross-language search will not be very helpful. First, the syntactic features of each language will skew the search. For example, since \JAVA{} is static typed, variables are declared with a datatype, while variables in \PYTHON{} lack these tokens due to dynamic typing in \PYTHON{}. Second, even if the language-specific keywords are ignored, there is an over reliance on the names of variables and libraries to infer behavioral context which may not always succeed. For example, ~\sfref{javaLoop} uses \mintJ{evens} to denote a \JAVA{} list, while ~\sfref{pyComp} uses \mintJ{nums} to represent the same \PYTHON{} array. Still, identifier names can be informative in describing the behavior of code~\cite{xin2017leveraging, xin2019revisiting} and are thereby useful as a metric.

\input{figs_tokens_and_trees}

Using an AST-based approach to identify similar code across languages is useful but not consistently viable due to the language-specific constructs. For example, ~\sfref{javaLoop} uses a standard \mintJ{for} loop to populate the list while ~\sfref{pyComp} uses \mintP{list-comprehension} which is a pythonic construct for the same task. The nearest structural match would be ~\sfref{pyOdds} since it also uses a \mintJ{for} loop. However, the functions ~\sfref{javaLoop} and ~\sfref{pyOdds} are behaviorally different. In such cases, a dynamic approach based on IO similarity would reveal the differences.

Behavioral approaches also have their limitations. For example, \mintJ{IntStream} from ~\sfref{javaStream} is specific to \JAVA{} v8 and above. Similarly ~\mintP{xrange} from ~\sfref{pyComp} is specific to \PYTHON{} v2.x.  Hence, the right version of the language and libraries is a prerequisite and in many cases a major bottleneck for dynamic similarity.

There is no single best similarity measure for cross-language code-to-code search. It depends on multiple varying criteria which cannot be generalized for all cases. Hence, we need a code-to-code search technique that enables search based on multiple similarity measures. \added{This can be handled by aggregating the multiple measures into a single measure or by using all the measures in tandem.}

When aggregating into a single measure, it is easier to compare but there are limitations. 
Search methods~\cite{kirkpatrick1983optimization, kennedy1995particle} and evolutionary algorithms~\cite{fonseca1993genetic} convert multiple search objectives to a single-objective search problem~\cite{miettinen2012nonlinear}. 
Using such methods is not very optimal~\cite{zitzler1998evolutionary} as ranking the results would be very subjective if the objectives are independent \added{or weakly correlated to each other}. An aggregated approach can also lead to bias in comparison~\cite{clark1976effects}.
This motivates the use of an approach to ranking that preserves each of the similarity measures, using them in tandem. 
\added{In this section, we have shown that code that matches on one measure (e.g.,~AST similarity would match ~\sfref{javaLoop} with ~\sfref{pyOdds}) may differ on another measure (e.g., behavioral similarity shows ~\sfref{javaLoop} and ~\sfref{pyOdds} are different). When aggregating into a single measure, such nuances are often  lost. }

Non-dominated ranking orders search results across multiple independent search objectives without aggregating them.
\added{We select three similarity measures to represent the context, structure and behavior across programming languages. These measures are weakly and moderately correlated to each other, presenting non-overlapping perspectives when used to compare code. } 
Further, non-dominated ranking could provide information sufficient to explain why one result is ranked above another in a meaningful way, an ability that is lacking in aggregated approaches (e.g., \textit{"~\sfref{javaLoop} and ~\sfref{pyOdds} have more structural similarity"}). While we do not investigate the value of the explanations in this work, we conjecture that such explanations may be useful when developers discern between search results and leave that for future exploration.


\section{\longapproach{}}
\label{sect:longapproach}



\fref{mocs} depicts the general workflow of \approach{}:

\begin{enumerate}
    \item Offline, a \textit{Repository} is crawled to extract \textit{Code Snippets} (e.g., GitHub, a local File System.)
    \item  Offline, Indices are created for each of the following:
    
    \begin{enumerate}
        \item A \emph{Token Index} for code names and libraries (\sect{contextSearch}).
        \item A language-agnostic \textit{AST Index} for code structures (\sect{syntacticSearch}).
        \item If the code can be executed, the \textit{IO Index} is recorded (\sect{semanticSearch}).
    \end{enumerate}
    \item During search, a \emph{Code Query} is processed in the same manner as Steps 2(a)-(c), to gather \emph{Tokens}, \emph{AST}, and \emph{IO} information.
    \item \textit{Non-Dominated Sorting} identifies \textit{Search Results}, which are ranked and returned to the user (\sect{multiSearch})
\end{enumerate}

\noindent We illustrate \approach{} using the code examples from \fref{codeSamples}.

\input{figs_mocs}


\subsection{Token-based Search}
\label{sect:contextSearch}


Fragments of code that are contextually similar often use similar variable names~\cite{relf2004achieving}, though the naming conventions vary by language. For example, \JAVA{} primarily uses \mintJ{camelCase} conventions while \PYTHON{} uses \mintP{snake_case}.
Libraries across languages tend to share similar function names or contexts~\cite{allamanis2014learning}; for example \mintinline{Java}{List} class from \mintinline{Java}{java.util} library and \mintinline{Python}{list} from the \PYTHON{} \mintP{__builtin__} library are both commonly used to represent an array.
Developers tend to describe the code in comments based on the functionality~\cite{hindle2012naturalness}. We infer context by extracting non-language-specific tokens from source code and comments as follows:
\begin{enumerate}
    \item Remove language-specific keywords based on the  documentation~\cite{javaKeywords,pyKeywords}. For example, \JAVA{} tokens \mintJ{public} and \mintJ{static}, and \PYTHON{} tokens  \mintP{def} and \mintP{assert}, are all removed. 
    \item Remove frequently-used words used in a language based on  common coding conventions. For example, in \PYTHON{}, the token \mintP{self} is often used to denote the class object.
    \footnote{Complete lists of the removed tokens are available~\cite{mocs}.} 
    \item Remove common stopwords from the English vocabulary~\cite{bird2009accessing}, such as \mintJ{does} and \mintJ{from}.
    \item Split tokens to address language-specific nomenclature. Variables typically use \emph{camelCase} in \JAVA{} and  \emph{snake\_case} in \PYTHON{}. These are split into \{``camel'' and ``case''\}  and \{``snake'' and ``case''\}, respectively.
    \item Remove tokens of length less than \mintJ{MIN_TOK_LEN}. 
    \item Convert all the tokens to lower case.
\end{enumerate}

A repository of code is tokenized using the above approach and stored in an ElasticSearch~\cite{gormley2015elasticsearch} index. For a user's code query, the tokens generated from the indexing approach are looked up in the search index and the best matched results are returned using the \emph{token similarity distance} (\dToken{}). This distance is the same as the Jaccard Coefficient~\cite{niwattanakul2013using}  and is defined as follows:
\[ d_{token} = \frac{|tokens_{query} \cap tokens_{result}|}{|tokens_{query} \cup tokens_{result}|} \]

\noindent \dToken{} will range from $[0.0, 1.0]$. Larger values of \dToken{} indicate higher similarity between the query and result.

For the functions in \fref{codeSamples}, the generated tokens using this approach are shown in \fref{tokensAndTrees} and the token similarity distance for each pair of functions is shown in \tref{codeMetrics} (\dToken{}). 
If the \JAVA{} function \sfref{javaLoop} is the query, the best \PYTHON{} result would be \sfref{pyComp} (\dToken{} = 0.067).
The common token \mintJ{max} extracted from these functions help in identifying this similarity. 
Note that none of the functions in \fref{codeSamples} have code comments, while in our implementation the comments are analyzed. 
 

In many cases the token-based analysis cannot yield ideal results. 
It relies on self-describing snippets; the choice of variable names, function names, libraries used, and comments all impact the results. 
Not all code snippets adhere to intuitive naming conventions. 
For example, in \sfref{javaStream}, the programmer chose very generic names. Still, we show in \sect{rqSearchSvM} that our approach for tokenization yields more precise results compared to full text search. 

\input{tab_metrics_comp}

\subsection{AST-based Search}
\label{sect:syntacticSearch}

A tree-based representation for comparison across languages is challenging since there is no generic AST representation  that encompasses syntactic features of different languages. Traditional AST parsers like \mintJ{ANTLR}~\cite{parr2013definitive}, \mintJ{JavaParser}~\cite{javaParser}, \mintP{python-ast}~\cite{pythonAST} modules use different grammars to denote similar features. For example, a function node in \mintJ{JavaParser} is represented as \mintJ{MethodDeclaration} while the \mintP{python-ast} parser represents the node as \mintP{FunctionDef}. As a result, to compare ASTs of different languages  requires a mapping scheme between each pair of programming languages.

For better scalability to additional programming languages, we built a parser for a generic AST. By mapping the ASTs for \JAVA{} and for \PYTHON{} onto the generic AST, we can compare across these languages (see \sect{languageExtension} for a discussion on scalability). The generic AST is based on our intuition and chosen languages, and there may be more effective or efficient representations. 
It contains a superset of the language features, as follows:

\begin{itemize}
    \item \textbf{Common control structures}: Control structures are simplified and clustered. For example, the \mintJ{loop} node is used for the \JAVA{} constructs: \mintinline{Java}{for}, \mintJ{forEach}, \mintinline{Java}{while} and \mintinline{Java}{do-while}; and \PYTHON{} constructs: \mintP{for}, \mintP{while}, and \mintP{list-comprehension}. 
    
    \item \textbf{Normalizing Variable}: Variables are denoted as \mintJ{var} nodes.
    
    \item \textbf{Normalizing Literals}: Literals are denoted as \mintJ{lit} nodes. 
    
    \item \textbf{Normalizing Operators}: Operators are denoted as \mintJ{op} nodes. 
    
    \item \textbf{Language specific features}: If a feature is implemented in only one language, a custom node is created. For example, \mintJ{switch} is specific to \JAVA{} and not supported in \PYTHON{}. As a result, a custom node \mintJ{switch} is created for this statement.
\end{itemize}

Similarity between ASTs is computed using the Zhang-Sasha algorithm~\cite{zhang1989simple} (\dAST{}). The algorithm computes the minimum number of edits required to transform an ordered labeled tree to another ordered labeled tree in quadratic time.
\dAST{} will range from $[0, \infty)$. Lower values of \dAST{} are associated with higher similarity. 

For the functions in \fref{codeSamples}, the generated ASTs are shown in \fref{tokensAndTrees} and the AST edit distance for each pair of functions is shown in \tref{codeMetrics} (\dAST{}). Based on \dAST{}, a query with \PYTHON{} function \sfref{pyComp} yields \sfref{javaLoop} as the best \JAVA{} result (\dAST{} = 7). 
Notably, the syntactic constructs of the two functions are also different. The \PYTHON{} search query \sfref{pyComp} uses a \mintP{list-comprehension} which is a \PYTHON{} feature and the \JAVA{} search result \sfref{javaLoop} uses a \mintJ{for} loop. Identifying the matching search result is possible, since \mintP{list-comprehension} and the \mintJ{for} loop are denoted as \mintJ{loop} nodes in the grammar for the generic AST. 

There are cases where a generic AST-based approach is non-ideal. 
For example, if the \JAVA{} function \sfref{javaLoop} is queried using \dAST{}, the best \PYTHON{} result would be \sfref{pyOdds}. This is because both functions use traditional \mintJ{for} loops and updates the return array sequentially, and yet, the search result is behaviorally different from the query. Such scenarios can be handled using dynamic similarity.


\subsection{Input-Output based Search}
\label{sect:semanticSearch}

Dynamic search in \approach{}, is performed by clustering code based on the their IO relationship. 
To determine the IO relationship between two pieces of code, we use \slacc{}~\cite{mathew2020slacc}, a publicly-available IO-based cross-language code clone detection tool. 
SLACC segments code into executable snippets of size greater than \mintJ{MIN_STMTS} and executed on \mintJ{ARGS_MAX} arguments generated using a grey-box strategy. The executed functions are then clustered using a similarity measure ($sim$) based on the inputs and outputs of the functions. Consider a query $q$ and a potential search result $s$. Let $Q$ and $S$ be sets of segments identified by \slacc{} from $q$ and $s$ respectively. We define the IO similarity as:

\[d_{IO} (q, s)  = \frac{1}{\lvert Q \rvert} \sum_{q_i \in Q} \underset{s_k \in S}{maximize} \; sim(q_i, s_k) \]

\noindent  The value \dIO{}  range from $[0.0, 1.0]$. 
Higher similarity corresponds to higher values of \dIO{}.

The IO similarity between any $q$ and $s$ is not commutative. 
This is because it is often preferred for a search result to contain extra behavior as compared to the query~\cite{stolee15Code}. Also, there may be a many-to-one mapping where multiple query segments match with a single segment in the result. 
Consider an example: let $Q = \{q_1, q_2, q_3, q_4, q_5\}$ be set of five segments and $S = \{s_1, s_2, s_3\}$ be set of three segments identified by \slacc{}. Segments $q_1$, $q_2$ and $q_3$ find segments $s_1$, $s_2$ and $s_1$ to be the most similar, respectively, with similarity scores ($sim$) of $sim(q_1, s_1) = 0.8$, $sim(q_2, s_2) = 0.95$ and $sim(q_3, s_1) = 0.7$. Notice how $s_1$ is identified as the closest match for both $q_1$ and $q_3$. 
Segments $q_4$ and $q_5$ did not find segments in $S$ with similarity greater than $0.0$. 
In this case, $d_{IO}(q,s) = \frac{0.8 + 0.95 + 0.7 + 0.0 + 0.0}{5} = 0.49$.  

As a practical example, say a developer is looking for a \JAVA{} API for \emph{QuickSelect}\footnote{from \mintJ{org.apache.datasketches}}, which finds the $k^{th}$ smallest number from an array of integers. 
It has a method that identifies a random pivot in the array and a method that swaps values. 
However, these methods do not call each other. 
Thus, to characterize the behavior of this file, we characterize and aggregate the behavior of segments of the file. 
Then, when comparing to a custom \PYTHON{} \emph{QuickSort}\footnote{\href{https://stackabuse.com/quicksort-in-python/}{stackabuse.com/quicksort-in-python}} API that has a function to recursively find a random pivot and perform a swap operation, a match is identified even though the number of methods and how they accomplish the same task are different.

The dynamic similarity between all the functions in ~\fref{codeSamples} is shown in ~\tref{codeMetrics}. We noticed in \sect{syntacticSearch} that ~\sfref{javaLoop} and ~\sfref{pyOdds} were very similar based on \dAST{}, but functionally different. 
Using behavioral analysis, we see that they are indeed functionally different as $d_{IO} = 0.5$ for both measures of $d_{IO}$. In contrast, \sfref{javaLoop} and \sfref{javaStream} are similar based on  behavior ($d_{IO} = 1.0$), even though structural similarity ($d_{AST}$) is low.


\subsection{Non-Dominated Ranking}
\label{sect:multiSearch}





We use Non-Dominated Sorting, which is a component of NSGA-II~\cite{deb2002fast}, and orders results with multiple objectives without aggregation.
\approach{} uses this algorithm to rank search results based on \dToken{}, \dAST{}, and \dIO{}. \added{We note that the similarity measures considered in this work are weakly correlated, as described in \sect{nonDomImpact}, making this an appropriate algorithmic choice. }
We use the algorithm in a novel context; this is the first work that uses non-dominated sorting for code-to-code search.


In our context, each similarity measure is an objective. \approach{} incorporates non-dominated ranking as follows:
\begin{enumerate}
    \item \textbf{Individual Search}: For a query, top \mintJ{TOP_N} search results are fetched
    using each similarity measure (\dToken{}, \dAST{} and \dIO{}) independently.
    \item \textbf{Merge}: The individual search results are merged \added{such that duplicate instances of search results are removed. 
    } 
    \item \textbf{Sort}: The merged results are sorted by NSGA-II~\cite{deb2002fast} by measuring the \emph{dominance} of one result over the other.  
\end{enumerate}

\noindent A search result $s$ is said to dominate a search result $t$, if $s$ is no worse than $t$ in any objective and is better than $t$ in at least one objective. Otherwise, there is a tie. In case of a tie, we select the result that has the dominant objective closest to the optimal value. \added{For example, consider the following scenarios of the relationship between $s$ and $t$ and three hypothetical similarity measures, $d_A$, $d_B$, and $d_C$, where higher values mean higher similarity:}

\begin{center}
\begin{tabular}{r c c c  l}
\textbf{scenario} & $\bm{d_{A}}$ & $\bm{d_{B}}$ & $\bm{d_{C}}$ & \textbf{winner} \\ \midrule
1 & $\textbf{s} > t$ & $\textbf{s} > t$ & $\textbf{s} > t$ & $s$\\
2 & $s = t$ & $s = t$ & $\textbf{s} > t$ & $s$\\
3 & $s = t$ & $s < \textbf{t}$ & $\textbf{s} > t$ & tie\\
4 & $s < \textbf{t}$ & $s < \textbf{t}$ & $\textbf{s} > t$ & tie\\
\end{tabular}
\end{center}

\noindent \added{For scenario 1, $s$ is better than $t$ on all measures, making $s$ the winner. In scenario 2, since $s$ is better than $t$ on one measure, and is never worse than $t$, $s$ is the winner. In the third scenario, $s$ is better than $t$ on one measure ($d_C$), and worse on another ($d_B$). Therefore, there is a tie. Similarly on scenario 4, $s$ is worse than $t$ on two measures and better on one, so it is also a tie. Ties are broken by looking at the search results that are better for each  similarity measure and then comparing to optimal values (typically $1$ or $0$, depending on whether high or low values represent better similarity). }

\added{Using the examples from \sect{motivation}, consider the \PYTHON{} functions ~\sfref{pyComp} and ~\sfref{pyOdds} in \fref{codeSamples} as queries. The potential cross-language results are \sfref{javaLoop} and \sfref{javaStream}. We show the relationships between the potential results using three similarity measures; see \tref{codeMetrics} for specific values. The winner for each comparison is bolded for clarity.}

\begin{center}
\begin{tabular}{r c c c l}
\textbf{query} & $\bm{d_{tokens}}(\uparrow)$ & $\bm{d_{AST}}(\downarrow)$ & $\bm{d_{IO}}(\uparrow)$ & \textbf{winner} \\ \midrule
\sfref{pyComp} & \textbf{\sfref{javaLoop}} > ~\sfref{javaStream}   & \textbf{\sfref{javaLoop}} < ~\sfref{javaStream}    & \sfref{javaLoop} = ~\sfref{javaStream} & \sfref{javaLoop}\\
\sfref{pyOdds} & \sfref{javaLoop} = ~\sfref{javaStream}            & \textbf{\sfref{javaLoop}} < ~\sfref{javaStream}    & \sfref{javaLoop} < \textbf{~\sfref{javaStream}} & ~tie\\
\end{tabular}\\
\end{center}

\added{When the query is \sfref{pyComp},  \sfref{javaLoop} is better than ~\sfref{javaStream} on two of the measures, and equal on the third, thus making \sfref{javaLoop} the winner. 
When the query is \sfref{pyOdds}, \sfref{javaLoop} is the winner for $d_{AST}$ and ~\sfref{javaStream} is the winner for $d_{IO}$, meaning we need to break the tie. }


\added{To break ties, we compute distances between each search result and the optimal value for each similarity measure (omitting similarity measures on which the results are tied).
The optimal value for $d_{AST}$ is $0$, as that represents isomorphic ASTs. 
The optimal value for \dIO{} is $1$, as that represents a perfect match in code behavior (i.e., the search result and query return the same output for all the provided inputs). 
The optimal value for \dToken{} is also $1$, as this represents highly similar syntax.
We use a normalized distance because the similarity measures have different ranges of values. Thus, normalizing ensures a uniform comparison scale between the different similarity measures and subsequently avoid the precedence of one similarity measure over other similarity measures.
The normalized distance of a similarity measure ($X$) on a snippet $s$ is computed as $\frac{d_{X}(s) - min(d_{X})}{max(d_{X}) - min(d_{X})}$. For \dToken{} and \dIO{}, the $max$ and $min$ values are $0.0$ and $1.0$ respectively. In the case of \dAST{}, the $min$ value is $0$ and the max value is set to the largest value of \dAST{} from all the individual search results. 
This is because \dAST{}, can theoretically be infinitely large so we use the largest observed value. For example,  $max(d_{AST})$ for the query \sfref{pyComp} is $21$ from \tref{codeMetrics}. }

\added{Continuing with the example, the normalized distance for ~\sfref{javaLoop} to the optimal $d_{AST}$ is $0.048$. We do not need to consider the distance for ~\sfref{javaStream} since ~\sfref{javaLoop} was the winner for $d_{AST}$. }
\added{
The normalized distance between \dIO{} of ~\sfref{javaStream} is $0.5$.}
\added{ Since the normalized \dAST{} of \sfref{javaLoop} is closer to the optimal value compared to the normalized \dIO{} of ~\sfref{javaStream}, \approach{} ranks \sfref{javaLoop} as the winner for the query ~\sfref{pyOdds}.}

We note that similarity measures characterizing other code relationships, such as software metrics~\cite{bansal2014selecting,patenaude1999extending}, could be added with relative ease. Non-domination ranking preserves each objective's independence and there are no weights that require tuning; see ~\sect{similarityExtension}.

\input{tab_rqs}

\section{Research Questions}
\label{sect:rq}
There does not exist a cross-language code-to-code search tool to compare against directly (see \sect{related}). Thus, our evaluation assesses each part of \approach{}: the ranking algorithm, within-language code-to-code search compared to state-of-the-practice and state-of-the-art tools, and cross-language clone detection. The first research question (RQ) examines the similarity measures and ranking:

\begin{resQues}
     Does non-dominated ranking using tokens, AST and IO yield better results for cross-language code-to-code search as compared to any subset or aggregation of those search similarity measures?
\end{resQues}

After validating the choice of using multiple code similarity measures and non-domination ranking, \approach{} is compared to the state-of-the-practice search in GitHub Search and ElasticSearch which are based on full text search. We ask:
\begin{resQues}
    How effective is  \approach{} in cross-language code-to-code search compared to state-of-the-practice public code search tools? 
\end{resQues}

\facoy{}~\cite{kim2018facoy}, the state-of-the-art in code-to-code search is within-language, but \approach{} is a multi-language tool. We limit our tool to within-language code-to-code search and evaluate it against \facoy{}.


\begin{resQues}
    How effective is \approach{} in within-language code-to-code search as compared to the state-of-the-art?
\end{resQues}

Code-to-code search is often used in clone detection~\cite{ragkhitwetsagul2019siamese,reiss2009semantics}. 
Using \approach{} for clone detection, we compare against \astLearner{}~\cite{perez2019cross}, \clcdsa{}~\cite{nafi2019clcdsa}, and \slacc{}~\cite{mathew2020slacc}:
\begin{resQues}
    Can \approach{} effectively detect cross-language code clones?
\end{resQues}


\section{Study}
The setup for each RQ is different.
In all evaluations we make a best effort to be fair in the comparison. 
The RQs are summarized in \tref{rqDesc}, which lists the application (either code-to-code search or clone detection), baseline approaches, language(s), and benchmarks.

\subsection{Data}
The data used in this study are available online~\cite{mocs}.

\subsubsection{AtCoder (AtC)}
We require a labeled set of similar code snippets in multiple programming languages for queries and search results. Hence, like prior studies~\cite{perez2019cross, nafi2019clcdsa}, we use AtCoder~\cite{atCoder} to create a dataset of similar code snippets across different programming languages.
Competitive programming contests like AtCoder~\cite{atCoder} have open problems where users can submit their solutions in most common programming languages. Solutions which are syntactically incorrect or do not pass the extensive test suite are filtered out by AtCoder. All the accepted solutions for a single problem implement the same functionality and are behavioral code clones. If a search query and a result belong to the same problem, we consider the result to be valid and the query-result pair as valid code clones; the problem solutions are the ground truth in our experiments. We limit our study to the most recent 398 problems which had solutions in \JAVA{} or \PYTHON{}. For these problems, we crawled 43,146 files from all the \emph{accepted} \JAVA{} and \PYTHON{} solutions. \tref{data} lists an overview of the dataset used for the study; 307 of the 398 problems have both a \JAVA{} and \PYTHON{} solutions.

\subsubsection{BigCloneBench (BCB)}
BigCloneBench~\cite{svajlenko2015evaluating} is one of the largest publicly available code clone benchmarks for \JAVA{} with over 55,000 source code files harnessed from approximately 25,000 open-source repositories. \tref{data} lists a summary of BigCloneBench. We consider query-result snippets belonging to the same functionality as a valid search result. Fragments of code with less than 6 lines or 50 tokens are not considered which is a standard minimum clone size for benchmarking~\cite{bellon2007comparison, kim2018facoy, svajlenko2015evaluating}.




\subsection{Baselines}
We compare \approach{} to each of the other tools by searching over the same data sets. For RQ3 and RQ4, we used the source code in the GitHub repositories of the tools for experimentation.

\subsubsection{RQ2 -- Text Search}
\label{sect:elasticSearch}
Google search is commonly used by developers for code search~\cite{stolee2014solving}. 
Textual queries can take the form of keywords, expected code, or exceptions raised. 
In our study, Google failed to index our code repository after a six week wait. As a result, we turned to a custom full text search using ElasticSearch~\cite{gormley2015elasticsearch} which takes in a code snippet, tokenizes the code and identifies results based on Lucene's Practical Scoring Engine~\cite{azzopardi2017lucene4ir}. 
For this study, each \JAVA{} and \PYTHON{} file is added to an ElasticSearch index and searched using the ElasticSearch programmatic search API.

\subsubsection{RQ2 -- GitHub Search}
\label{sect:githubSearch}
GitHub search engine is an IR-based search model over code repositories, including issues, pull request, documentation, and code data~\cite{githubSearch}. 
Using the built-in code search on GitHub, code can be searched globally across all of GitHub, or searched within a particular repository or organization. 
We add the \JAVA{} and \PYTHON{} files from the dataset to a single GitHub repository and search within the repository using the GitHub Search API~\cite{githubAPI}. 

\subsubsection{RQ3 -- \facoy{}}
\facoy{}~\cite{kim2018facoy} is a \JAVA{}-based code-to-code search tool that uses a query alternation approach using relevant keywords from StackOverflow Q\&A posts.
\facoy{} can be modified to change its search database from Q\&A posts to custom datasets. In our experiments, we redirected the search to the repositories of code from the AtCoder and BigCloneBench datasets. Similar to the experiments in the \facoy{} evaluation when comparing against research tools, 
 \facoy{} does not use StackOverflow in our baseline. 



\subsubsection{RQ4 -- \astLearner{}}
Perez and Chiba developed a semi- \linebreak supervised cross-language syntactic clone detection method that we call \astLearner{}~\cite{perez2019cross}. It uses a skip-gram model and an LSTM based encoder. The encodings  train a  feed forward neural network classifier using negative sampling to identify clones.
\astLearner{} considered code as clones if the classifier score is greater than $0.5$.

\subsubsection{RQ4 -- \clcdsa{}}
Cross Language Code Clone Detection~\cite{nafi2019clcdsa}   (\clcdsa{}),  uses syntactic features and API documentation to detect cross-language clones in \JAVA{}, \PYTHON{} and C\#. 
Nine features are extracted from the AST; API call similarity is learned using API documentation and a Word2Vec~\cite{mikolov2013efficient} model. 
The vectorized features train a reconfigured Siamese architecture~\cite{baldi1993neural} using a large amount of labeled data. 
\clcdsa{} uses cosine similarity to detect clones; the best F1 scores were when the similarity threshold was $0.5$. 

\subsubsection{RQ4 -- \slacc{}}
Simion-based Language-Agnostic Code \linebreak Clones~\cite{mathew2020slacc} (\slacc{}), uses IO behavior to identify clones. It is also used in \approach{} for dynamic similarity. Here, we use \slacc{} as a baseline in its original context, clone detection. We use the same values for the hyper-parameters set by the authors of \slacc{}: \mintJ{MIN_STMTS} is set to $1$; \mintJ{ARGS_MAX} is set to $256$; \mintJ{SIM_T} is set to $1.0$. 

\input{tab_data}

\subsection{Metrics}

\subsubsection{Code Search}
\label{sect:metrics:search}
For code search applications (RQ1, RQ2, RQ3), we use \emph{Precision@k}, \emph{SuccessRate@k}, and \emph{MRR}. 

\emph{Precision@k} or \emph{P@k} is the average percentage of relevant results in the top-k search results for a query~\cite{gu2018deep,kim2018facoy}. 
\emph{SuccessRate@k}  or \emph{SR@k} is the percentage of queries for which one or more relevant result exists among the top-k search results~\cite{luan2019aroma,gu2018deep}.
\emph{MRR} is the {M}ean {R}eciprocal {R}ank of the relevant results for a query~\cite{kim2018facoy,luan2019aroma,gu2018deep}. 

Consider a query $q$ in a set of queries $Q$. $R_q^k$ is set of all relevant results in the top k results for $q$. $BR(q)$ is the rank of the first relevant search result for $q$. $\delta_k$ is an indicator function which returns 1 if the input is less than or equal to $k$ and 0 otherwise. Mathematically,
\begin{small}
\[{P@k = \frac{\sum\limits_{q\in Q}{\frac{|R_{q}^{k}|}{k}}}{|Q|} \quad SR@k = \frac{\sum\limits_{q\in Q}{\delta_k(BR(q))}}{|Q|} \quad MRR = \frac{\sum\limits_{q\in Q}{\frac{1}{BR(q)}}}{|Q|}}\]
\end{small}

\noindent \emph{Precision@k}, \textit{SuccessRate@k} and \textit{MRR} range $[0.0, 1.0]$. For better readability, in the rest of study, we report these metrics as percentages ranging between $[0, 100]$. For $k=1$, \textit{Precision@k} and \textit{SuccessRate@k} are the same.  For higher values of $k$, \textit{SuccessRate@k} indicates whether there is something relevant in the results, \textit{Precision@k} measures how relevant the $k$ results are on average. We set $k=\{1,3,5,10\}$. Higher values of \textit{MRR} imply relevant results are ranked higher in the results.

\subsubsection{Clone Detection}
\label{sect:metrics:clone}



For clone detection~\cite{perez2019cross,nafi2019clcdsa} (RQ4), we use \emph{Precision}, \emph{Recall} and \emph{F1} score.
\emph{Precision} (\emph{P}) is the ratio of valid clones to the number of retrieved clones.
\emph{Recall} (\emph{R}) is the ratio of the number of accurately detected clones to the number of total actual clones.   
\emph{F1} or \emph{F-Measure}, is the harmonic mean of precision and recall. We define  $\lvert C_+ \rvert$ as the number of  valid clones identified, $\lvert C_- \rvert$ as the number of valid clones not identified, and $\lvert NC_+ \rvert$ as the number of invalid clones identified:
\begin{equation*}
    \tiny
  P = \frac{|C_+|}{|C_+| + |NC_+|} \quad\quad R = \frac{|C_+|}{|C_+| + |C_-|} \quad\quad F1 = \frac{2*P*R}{P + R}  
\end{equation*}

\noindent 
\textit{Precision}, \textit{Recall} and \textit{F1}  range from $[0.0, 1.0]$. Like the code search metrics, we report \textit{Precision}, \textit{Recall} and \textit{F1} as percentages between $[0, 100]$ for better readability. Higher values of \textit{precision} mean the detected clones contain fewer false positives and higher values of recall mean more clones were identified with fewer false negatives.


\subsection{Experimental Setup}
\label{sect:expSetup}
Our experiments were run on a Ubuntu 18.04 LTS Virtual Machine with 32 CPUs and 64GB memory using a Dell PowerEdge R640 server with Intel Xeon Silver 4210 CPU @ 2.2 GHz and VMware ESXi 6.7.0 hypervisior.
The experiments have four hyper-parameters:

\subsubsection{Minimum token size} (\mintJ{MIN_TOK_SIZE} in \sect{contextSearch}) This is set to three. IR based techniques~\cite{hindle2012naturalness, xin2017leveraging} on source code find that tokens less than three characters are irrelevant.
\subsubsection{Minimum segment size} (\mintJ{MIN_STMTS} in \sect{semanticSearch}) A small value of \mintJ{MIN_STMTS} results in more granular snippets. We set it to $1$ for maximum number of behavioral snippets of code.
\subsubsection{Maximum number of arguments} (\mintJ{ARGS_MAX} in \sect{semanticSearch}) Prior work finds \mintJ{ARGS_MAX}=256 was sufficient for cross-language clones in Google Code Jam~\cite{googleCodeJam} , so we use the same. 
    
\subsubsection{Number of individual search results} (\mintJ{TOP_N} in \sect{multiSearch}) This is set to $100$. We experimented on ~\approach{} with $10\%$ of the AtCoder dataset varying \mintJ{TOP_N} in $\{10,20,50,100,200,500\}$. For \mintJ{TOP_N} greater than $100$, we see a minimal change in the code search metrics.  Hence, for each individual search, we fetch the top $100$ search results.

\section{Results}
\label{sect:results}

We present the results of each RQ in turn. 

\subsection{RQ1: Single vs Multiple Search Similarity Measures}
\label{sect:rqSearchSvM}

In a cross-language search context,  we compare the results of \approach{} with  multiple search similarity measures to \approach{} with subsets of the similarity (e.g., \approachAST{} is \approach{} with only the AST similarity). 
The validation of this study was performed using `leave-one-out' cross-validation~\cite{sammut2010leave}  where each code fragment is used as a query against all other  fragments in the repository. We use this approach over the traditional k-fold cross validation since  `leave-one-out' is approximately unbiased and more thorough~\cite{luntz1969estimation}. 

Each of the 43,146 code fragments is used as a query. 
The results are detailed in \tref{crossLangSearch}. Overall, \approach{} outperforms the other formulations that use subsets of the similarity measures. It also outperforms an alternate ranking approach based on weighted measures (KDTree~\cite{bentley1975multidimensional}).

We observe that token-based search (\approach{}$_{tokens}$) and AST-based search (\approach{}$_{AST}$) are less precise individually compared to dynamic search (\approach{}$_{\slacc{}}$), but have higher success rate for $k=\{5,10\}$. 
When both the static similarity measures are used as parts of a bi-similarity search (\approach{}$_{static}$), we see better metrics compared to each similarity individually, and better metrics than the dynamic approach \approachSLACC{} in \textit{P@k} and \textit{SR@k} when $k > 1$. 

The power of the technique comes from using static and dynamic information without converting them into a single search metric. Rather than non-dominated ranking, an alternate avenue would be a weighted approach. For example, \weightedApproach{} uses \dToken{}, \dAST{} and \dIO{} to build a KDTree~\cite{bentley1975multidimensional}, a common approach used for information retrieval~\cite{deng1998omega, greenspan2003approximate}. 
Although \weightedApproach{} and \approach{}  use the same similarity measures for code search, the former under-performs on all metrics compared to the latter. This suggests that aggregation of similarity measures into a single measure does not help code search as these measures complement each other. 

\begin{slimbox}
Using non-dominated ranking with static and dynamic similarity measures improves the quality of results for code-to-code search compared to subsets or a weighted aggregation of measures.
\end{slimbox}

\input{tab_results_cross_language}

\subsection{RQ2: State-of-the-Practice Cross-language Code-to-Code Search}
\label{sect:rqSearch}
We compare \approach{} against GitHub Search (\sect{githubSearch}) and a custom full text search based on ElasticSearch (\sect{elasticSearch}). We use `leave-one-out' cross-validation with each of the 43,146 code fragments as a query. Results are shown in ~\tref{crossLangSearch}. 

We observe that between the textual code search tools, GitHub Search has better \textit{MRR}, \textit{Precision@k} and \textit{SuccessRate@k} compared to ElasticSearch except for \textit{SuccessRate@10}. Yet, GitHub Search and ElasticSearch are  worse off compared to \approach{} in all metrics. 

\begin{slimbox}
\approach{} obtains better \textit{Precision@k}, \textit{SuccessRate@k} and \textit{MRR} compared to GitHub Search and ElasticSearch. 
\end{slimbox}

\subsection{RQ3: State-of-the-Art Code-to-Code Search}
\label{sect:rqSearchSingleLang}
\input{tab_results_atc_bcb}

\facoy{}~\cite{kim2018facoy} is a state-of-the-art code-to-code search tool for \JAVA{}. 
Hence, we compare \approach{} against \facoy{} using \JAVA{} code snippets only. This reduces the AtCoder dataset to 351 problems with 20,673 \JAVA{} files. To ensure that the dataset is not skewed due to outlier projects with limited submissions, we use \JAVA{} projects with 10 or more submissions. Like RQ1 and RQ2, we use `leave-one-out' cross-validation with each of the 20,673 code fragments as a query and the remaining problems as the search index.

The results for \textit{MRR}, \textit{Precision@k} and \textit{SuccessRate@k} are tabulated in \tref{sameLangSearch}. 
\approach{} has better scores on all metrics compared to \facoy{}. Even if \approach{} is used with only static similarity measures (\approachStatic{}), it consistently performs better than \facoy{}. 

Since, \facoy{} supports only \JAVA{}, we also compare \facoy{} to \approach{} using BigCloneBench. This experiment  moves us toward evaluating the feasibility of \approach{} with open-source projects. We again use `leave-one-out` cross-validation where each file from BigCloneBench is used as a query and the other files are used as search results. A search result is considered valid if it has the same functionality group as the search query. 

Compared to AtCoder, the BigCloneBench dataset yields better results for all techniques. 
This is because the 43 functionalities in BigCloneBench have minimal overlap. 
This can be corroborated by the better scores for token-based search compared to the AST-based search on BigCloneBench dataset. In contrast, on AtCoder, AST-based search out-performs token-based search.
Like the AtCoder dataset, search based on a combination of measures (\approach{}$_{static}$, \approach{}) yield better results compared to \facoy{}.

Only 4,984 ($9\%$) of the files from BigCloneBench are executable by \slacc{};  the remaining files depend on external libraries. Thus, dynamic similarity (\approach{}$_{SLACC}$) has much lower scores  in ~\tref{sameLangSearch}. Subsequently, the inclusion of dynamic similarity hardly contributes to the results of \approach{} as highlighted by their similar values for \approach{}$_{static}$ and \approach{}. We dive deeper into the role of dynamic similarity in \sect{dynamicanalysisbenefit}.

\begin{slimbox}
Compared to state-of-the-art \JAVA{} code-to-code search \facoy{}, using dynamic information helps \approach{} obtains better search results when executable code snippets are present. In the absence of dynamic information, a combination of AST and token-based similarity measures still yields better results than \facoy{}.
\end{slimbox}


\subsection{RQ4: Cross-language code clone detection}
\label{sect:rqClone}

\input{tab_results_clones}

As there is no existing tool for cross-language code-to-code search, we instead compare to cross-language code clone detection techniques: \astLearner{}, \clcdsa{} and \slacc{}. While code-to-code search can be part of clone detection, they are different. For a given code snippet, code clone detection returns an identical code snippet and code-to-code search returns a set of potentially relevant snippets. {Hence, to use \approach{} as a clone detection tool, we select the top-1 ranked search result returned by non-dominated ranking.}

\astLearner{} and \clcdsa{} build deep learning models and require a training, validation and testing set. 
Hence we randomly divide our dataset into these three sets using the same approach adopted in \clcdsa{}~\cite{nafi2019clcdsa}. 
We only consider projects with at least 20 \JAVA{} and 20 \PYTHON{} submissions, reducing the dataset to 302 different problems. 
For each problem, we select ten submissions each from \JAVA{} and \PYTHON{} as part of the training set, five for the validation set and five for the test set. 
We used the default hyper-parameters from \astLearner{} and \clcdsa{} to build their models. 
Since \approach{} and \slacc{} do not use machine learning models, we add all the submissions from the training set to the search database and use the test set for evaluation. {We do not include the validation set in the search database to ensure a fair comparison to \astLearner{} and \clcdsa{}.} To account for variance, we repeat this step 10 times and report the mean \textit{precision}, \textit{recall} and \textit{F1} scores.

Results are shown in  \tref{crossLangClone}, \added{separating the techniques that use a single similarity measure (\emph{Single Sim.}) from those that use multiple similarity measures (\emph{Multi Sim.})}. \slacc{} is the most precise technique on this dataset but has extremely low \textit{recall} compared to other techniques, and hence the lowest F1. The low \textit{recall} on \slacc{} is because it requires executable code snippets. \approach{} has better \textit{precision} and \textit{recall} compared to the static similarity approaches \astLearner{} and \clcdsa{}. If \approach{} is used only with the static similarity measures (\approachStatic{}), the \textit{precision} and \textit{recall} is still better than \astLearner{} and comparable to \clcdsa{}. 



\begin{slimbox}
For code clone detection, \approach{} obtains better \textit{precision}, \textit{recall} and \textit{F1} scores compared to \astLearner{} and \clcdsa{}, without the need to build models. \approach{} has lower \textit{precision} to \slacc{} but much better \textit{recall} and \textit{F1} score.
\end{slimbox}

\section{Discussion}
We have evaluated \approach{} extensively against prior work in code-to-code search and clone detection. 
In all cases, it outperforms the competition without the need to build, train, or update models. 
In this section, we discuss the cost/benefit of dynamic analysis, the potential for scalability, and  threats to the validity.

\subsection{On the Cost/Benefit of Dynamic Analysis}
\label{sect:dynamicanalysisbenefit}
\input{tab_semantics_discussion}

In \sect{rqSearchSingleLang} and ~\tref{sameLangSearch}, we observe a low scores for code search using IO-based similarity (\approachSLACC{}) compared to other techniques due to the small sample of files in BigCloneBench ($9\%$) with executable code. To study the relative contribution of dynamic analysis to \approach{} results, we repeat the validation study on BigCloneBench but restricted to the files that can be executed (4,984).

Results on the executable dataset are slightly better for all the techniques compared to the complete BigCloneBench dataset (\tref{bcbSemantic}). 
Although \approachSLACC{} is slightly better than \approachStatic{}, executing snippets takes more time and memory, making code search slow and impractical if the runtime data are not cached. Since the gains are not very high with the BigCloneBench dataset, it might be sufficient to rely on static similarity in this case. 

However, this cannot be generalized across datasets. BigCloneBench is built on \JAVA{} code from open-source projects. For cross-language search (\tref{crossLangSearch}), using dynamic and static similarity measures vastly improves the results. This is due to the syntactic differences between languages which can be overcome in many cases with dynamic information~\cite{jiang2009automatic}. Hence, the benefit of including dynamic similarity data must be balanced against the cost and context.

\subsection{On Non-Dominated Sorting}
\label{sect:nonDomImpact}

\added{For cross-language code search, combining the search similarity measures using an aggregated weighted approach (\weightedApproach{}) results in lower MRR, \textit{P@k} and \textit{SR@k} compared to the non-dominated sorting approach (\tref{crossLangSearch}). As one potential explanation, this poorer performance for the aggregation approach could be a result of bias due to the independence or weak correlation between the three similarity measures~\cite{clark1976effects} . 
In this section, we explore the impact of the correlations between the similarity measures.} 


\input{tab_correlation}

\added{\tref{correlation} shows the Pearson's correlation ($r$) between the three similarity measures for cross-language and within-language snippets on $20$ repeats of $1000$ random pairs of snippets. 
Overall, for the cross-language analysis, we observe lower correlations compared to the within-language analyses. The cross-language correlations are weak ($0.20 \leq |r| \leq 0.39$)~\cite{moore2007basic}  to  moderate ($0.40 \leq |r| \leq 0.59$). 
The single-language correlations are moderate to strong ($0.60 \leq |r| \leq 0.79$).}

\added{Connecting this to our results, the weak to moderate correlations in the cross-language context may have contributed to relatively better performance of non-dominated sorting. 
Since non-dominated sorting is effective for search objectives with low correlation~\cite{zitzler1998evolutionary, tian2017effectiveness}, it seems appropriate for  cross-language code-to-code search.
Studies have also shown that non-dominated sorting works best for fewer objectives~\cite{elarbi2017new, zhang2007moea}. As \approach{} is extended with more metrics in the future, we will want to revisit this analysis. }

\added{However, as correlation impacts the performance of the ranking algorithm, non-dominated sorting is not a panacea. When the similarity measures are more strongly correlated, which our analysis shows is true for single-language code search, a different approach may be needed, such as aggregation or evolutionary algorithms. }

\input{figs_os_code_samples}

\subsection{Scalability Exploration}
We explore three scalability concerns: indexing and searching open-source code, adding new languages, and adding similarity measures.

\subsubsection{Open-Source Repositories}
\label{sect:ossScalability}

We used the AtCoder and BigCloneBench datasets  to benchmark our experiments, similar to prior art in  code search and clone detection~\cite{su2016code, su2016identifying, kim2018facoy, sajnani2016sourcerercc}. 
Yet, neither dataset is particularly realistic.  
AtCoder is composed of programming contest submissions and is not a true representation of open-source code. 
 BigCloneBench contains example code clones, making clone detection and code search relatively easier.
To some extent, these datasets set us (and the baselines) up for success. 

We want to explore how \approach{} could work with an arbitrary open-source project. 
To do this, we consider three popular open-source libraries for \JAVA{} and \PYTHON{}:  \mintJ{Guava} \JAVA{} library by Google,  \mintJ{commons-collections} \JAVA{} library by Apache Software Foundation, and \mintP{collections} \PYTHON{} 2.7 system library.




Consider the code snippets in ~\fref{osCodeSamples}. 
For this example, \approach{} uses  \sfref{guavaCountMultiSet} as the query, which counts the number of occurrences of an object in the \mintJ{MultiSet}. 
Across languages, \approach{} identifies a similar code snippet from the \mintP{collections} library in \PYTHON{}: \sfref{collectionsCount} returns the count of an element from a \mintP{Counter}. A \mintP{Counter} is a \PYTHON{} collection, like a bag, that takes elements and maintains a count of their occurrences. For this pair, we can see that they share few common tokens (\texttt{count, get}), do not have similar ASTs, but are behaviorally similar. Hence, the token-based and IO-based similarity in \approach{} influence the ranking of search results and returns \sfref{collectionsCount} as a valid search result for the query \sfref{guavaCountMultiSet}.

In  our experiments,  we see low scores for \approachSLACC{} since only around $9\%$  of the files in BigCloneBench had executable code. In this  open-source exploration, around $68\%$ of the \JAVA{} and all the \PYTHON{} classes had executable code. The presence of dependent code in the libraries compared to the isolated files in BigCloneBench actually facilitated more widely applicable behavioral analysis.

Thus, we conclude that \approach{} can be scaled to support open-source projects in the current implementation. The token-based and AST-based similarity measures for \approach{} can be used on any project or file(s) in its current version. Since the behavioral similarity measure used by \approach{} is heavily dependent on \slacc{}, scaling to support new projects would require the projects have all its dependencies satisfied and executable.

\subsubsection{Support for new languages}
\label{sect:languageExtension}
\approach{} currently supports \JAVA{} and \PYTHON{}.  While we have not demonstrated scalability to new languages, we comment on the effort required. 

For dynamic behavior, \approach{} is dependent on \slacc{}~\cite{mathew2020slacc}, so adding a new language to \approach{} requires support in \slacc{}. However, \approach{}$_{static}$ can be extended to new languages by adapting the token and AST analyses. A language-specific tokenizer like c-tokenizer~\cite{cTokenizer} or a generic tokenizer like ANTLR~\cite{parr2013definitive} can be used to parse code and convert it into tokens as detailed in \sect{contextSearch}. For the {AST}, \approach{} uses a generic AST to represent source code across different languages. Using a language-specific AST Parser like clang for C~\cite{lattner2018clang} or roslyn for .NET~\cite{roslyn}, code could be parsed and converted to the generic AST-based on the grammar available in the GitHub code repository for \approach{}~\cite{mocs}. If a feature specific to a language is not supported by the grammar, a new node should be created based on the feature's syntactic structure.

\subsubsection{Adding New Search Similarity Measures}
\label{sect:similarityExtension}
\approach{} uses three search similarity measures for code-to-code search, which provides a start for this line of research. New search similarity measures can be added or existing similarity measures can be replaced in \approach{}. First, a similarity measure to compare code snippets has to be defined. The similarity measure has to be a numerical value to support non-dominated ranking of the search results. Next, an index must be created characterizing the similarity measure. Lastly, the
similarity measure has to be updated in the
configuration file.

\subsection{Threats to Validity}
\label{sect:threats}
\hspace{0.3cm} \emph{Language Bias}. \approach{} was implemented for \JAVA{} and \PYTHON{} and may not generalize to other languages. 

\emph{Baseline Bias}. The ElasticSearch baseline for cross-language code-to-code search (in RQ2) is not an exact representation of a code-to-code search tool used by developers~\cite{sadowski2015developers}. 

\emph{Data Bias}. The datasets are from a programming contest and a code clone benchmark, which are not  representative of industrial or open-source coding practices. However, our initial investigation into open-source code (\sect{ossScalability}) revealed that \approach{} can be successful in that context, but more exploration is needed. 
\emph{Similarity Bias}. \approach{} uses three similarity measures based on syntactic and semantic features for code search based on the context, structure and IO behavior. Other similarity measures~\cite{bansal2014selecting, patenaude1999extending} are not explored in this study. But, \approach{} can be extended to support these similarity measures as described in Section ~\sect{similarityExtension}.

\section{Related Work}
\label{sect:related}
We present work in code similarity, search, and clone detection.

\subsection{Code Similarity}
Source code similarity is used to characterize the relationship between pieces of code in software engineering applications such as  program repair~\cite{gopinath2011specification, stolee2012toward, stolee2014solving, ke2015repairing, nguyen2013semfix}, code search~\cite{kim2018facoy,ragkhitwetsagul2019siamese,luan2019aroma}, software security~\cite{Ray:2013,Yue:2018,walenstein2007software} and identifying plagiarized code~\cite{baker1995finding}. 
 Code similarity can be measured through static or dynamic analyses.

Techniques that use static code attributes to compute similarity often parse code into an intermediate representation based on text~\cite{kamiya2002ccfinder,li2004cp,baker1995finding}, AST~\cite{baxter1998clone,jiang2007deckard} or graph-based~\cite{gabel2008scalable, li2012cbcd} and compute a measure for syntactic similarity. For cross-language syntactic similarity, most techniques are text-based~\cite{nguyen2017exploring,nafi2019clcdsa, kraft2008cross}. Tree- and graph-based approaches have not been explored for cross-language similarity due to language specific grammar. We tackle this challenge by creating a language-agnostic grammar by abstracting out common features across languages to build a generic AST (\sect{syntacticSearch})

Techniques that execute code to determine similarity are classified as dynamic. 
For some techniques, functions are adjudged to be similar if they have similar inputs, outputs, and side-effects~\cite{elva2012semantic,jiang2009automatic,su2016identifying,mathew2020slacc}. 
Other techniques use abstract program states after executions to analyze the behaviors of the code fragments~\cite{kim2011mecc,su2016code,perry2019semcluster}. Dynamic measures are particularly successful in detecting code clones across languages since it does not rely on syntactic properties~\cite{mathew2020slacc,jiang2009automatic}.
Limitations to this approach include the need to execute the code which dictates the granularity~\cite{deissenboeck2012challenges} and runtime.


\subsection{Code Search}
In code search, the goal is to find code that is similar to a given query.
Historically, developers have preferred general search engines such as \emph{Google} and \emph{Bing}  when searching for code to reuse~\cite{stolee2014solving,sim2011well,stolee15Code}. 
Some code search tools~\cite{krugler2013krugle,searchcode} use code snippets as the query, a problem called code-to-code search. 
Solutions to code-to-code search vary in several dimensions, we list three: within~\cite{gu2018deep,kim2018facoy} vs. across languages~\cite{nafi2019clcdsa, perez2019cross, luan2019aroma}, static~\cite{searchcode,kamiya2002ccfinder,jiang2007deckard} vs. dynamic analysis~\cite{mathew2020slacc,reiss2009semantics}, and index-based~\cite{kim2018facoy,luan2019aroma,githubSearch} vs. model based~\cite{gu2018deep,nafi2019clcdsa,perez2019cross}. 

In cross-language code-to-code search, the query is a code snippet in one source language and the results are from a different target language(s).
AROMA~\cite{luan2019aroma}, supports cross-language code-to-code search across \JAVA{}, Hack, JavaScript, and \PYTHON{} using static analysis based on the parse tree. Since AROMA is not publicly available, it is not used as a baseline in this study. 
\added{InferCode~\cite{bui2021infercode} is a self supervised cross-language (\JAVA{}, C, C++ and C\#) code representation approach using Tree-based Convolutional Neural Networks based on syntax subtrees. 
Since this work was performed in parallel to our study, we have not benchmarked \approach{} against InferCode and leave that for future work.}
\facoy{}~\cite{kim2018facoy} is a within-language code-to-code search tool on JAVA{} that uses query alteration to find semantically similar code snippets using Q\&A posts. 

\subsection{Clone Detection}
Clone detection is a special case of code-to-code search; results are identified as clones if they meet a specified similarity threshold.  
Clones are often categorized into four types: types I-III are based on syntax and type IV is based on behavior.

Most code clone detection tools~\cite{kamiya2002ccfinder, li2004cp, baxter1998clone, jiang2007deckard, gabel2008scalable, li2012cbcd, jiang2009automatic, su2016identifying} have been proposed for single language clone detection and on static typed languages like \JAVA{}~\cite{jiang2007deckard, koschke2006clone} and C~\cite{baxter1998clone, yang1991identifying,jiang2007deckard,kamiya2002ccfinder}. A small number of tools support cross-language code clone detection~\cite{mathew2020slacc,nguyen2017exploring,nafi2019clcdsa,perez2019cross}. 
API2Vec~\cite{nguyen2017exploring} detects clones between two syntactically similar languages by embedding source code into a vectors and subsequently comparing the similarity between the vectors. 
\clcdsa{}~\cite{nafi2019clcdsa} identifies nine features from the source code AST and uses a deep neural network to learn the features and detect cross language clones. 
Perez and Chiba~\cite{perez2019cross} propose an LSTM-based deep learning architecture using ASTs to detect clones in \JAVA{} and \PYTHON{} code. 
These three tools build machine learning models to detect code clones. As a result, these techniques require a large number of annotated training data to build the model and the hyper-parameters need to be carefully optimized to avoid over-fitting. 

SLACC~\cite{mathew2020slacc} is a cross-language code clone detection tool that uses IO profiles. It succeeds in detecting code clones with high precision between programming languages with different typing schemes. However, SLACC requires the code snippets to be executable and as a result has low recall and a large runtime. 

In a clone detection context, we use  \clcdsa{}, the Perez and Chiba approach, and \slacc{} as baselines for comparison (\cref{sect:rqClone}).

\section{Conclusion}
\label{sect:conclusion}
We present \approach{}, a cross-language code-to-code search tool that uses static and dynamic analyses. It uses two static similarity measures based on extracted tokens from source code and a tree edit distance based on a  generic AST, and one dynamic similarity measure to compute IO similarity. For a given code search query, these three similarity measures find results using non-dominated sorting. Our experimental evaluation on 98,645 \JAVA{} and \PYTHON{} files from AtCoder and BigCloneBench datasets show that \approach{} outperforms state-of-the-art code search tools \facoy{} and industrial benchmark of GitHub code search. 
We also compare \approach{} to state-of-the-art clone detection techniques using the AtCoder dataset and find that \approach{} has better \textit{Recall} and \textit{F1}. 
Cross-language code-to-code search appears to have a bright future, but more work is needed to evaluate it for more languages and in relevant applications.


\begin{acks}
We thank the anonymous reviewers for their valuable feedback. 
This work is supported in part by the National Science Foundation under NSF SHF \#\href{https://www.nsf.gov/awardsearch/showAward?AWD_ID=1645136}{1645136}, \#\href{https://www.nsf.gov/awardsearch/showAward?AWD_ID=1749936}{1749936}, and  \#\href{https://www.nsf.gov/awardsearch/showAward?AWD_ID=2006947}{2006947}. 
\end{acks}

\bibliographystyle{ACM-Reference-Format}
\bibliography{refs}


\end{document}

%% file: figs_code_samples_mini.tex
\captionsetup[subfigure]{labelfont=bf,textfont={bf},singlelinecheck=off,margin=12pt,skip=2pt}
\begin{figure}[!t]
    \begin{subfigure}{\textwidth}
        \centering
        \begin{minted}[xleftmargin=2em,linenos,firstnumber=1,fontsize=\footnotesize]{Java}
List<Integer> getEvens(int max) {
    List<Integer> evens = new ArrayList<>();
    for(int i = 0; i < max; i++)
        if (i % 2 == 0) 
            evens.add(i);
    return evens;
}
        \end{minted}
        \caption{\mintJ{Java}: \mintJ{for} loop to populate an array of even numbers}
        \label{fig:javaLoop}
    \end{subfigure}
    \par\bigskip
    \begin{subfigure}{\textwidth}
        \centering
        \begin{minted}[xleftmargin=2em,linenos,firstnumber=1,fontsize=\footnotesize]{Python}
def all_odds(n):
    odds = []
    n = range(n)
    for i in n:
        if i % 2 == 0: continue
        odds.append(i)
    return odds
        \end{minted}
        \caption{\mintJ{Python}: \mintJ{for} loop to populate an array of odd numbers}
        \label{fig:pyOdds}
    \end{subfigure}
    \par\bigskip
    \begin{subfigure}{\textwidth}
        \centering
        \begin{minted}[xleftmargin=2em,linenos,firstnumber=1,fontsize=\footnotesize]{Java}
Integer[] func(int x) {
    int[] n = IntStream.range(0, x).toArray();
    List<Integer> e = new ArrayList<>();
    for (int i=0; i<n.length(); i++)
        if (n.get(i) % 2 == 0)
            e.add(n.get(i));
    return e.toArray();
}
        \end{minted}
        \caption{\mintJ{Java}: \mintJ{List} of even numbers using version specific libraries}
        \label{fig:javaStream}
    \end{subfigure}
    \par\bigskip
    \begin{subfigure}{\textwidth}
        \centering
        \begin{minted}[xleftmargin=2em,linenos,firstnumber=1,fontsize=\footnotesize]{Python}
def even_nums(max_val):
    nums = xrange(max_val)
    return [i for i in nums if i % 2 == 0]
        \end{minted}
        \caption{\mintJ{Python}: \mintP{list} of even numbers using \mintP{list-comprehension}}
        \label{fig:pyComp}
    \end{subfigure}
    \caption{Different functions to return a filtered array of numbers implemented in \mintJ{Java} and \mintP{Python}. The code in (a), (c), and (d) are functionally identical. The code in (b) is different. }
    \label{fig:codeSamples}
\end{figure}

%% file: figs_tokens_and_trees.tex
\begin{figure}[tb]
    \centering
    \small

    \begin{tabular}{p{0.15in}m{1.75in}m{1in}}
        \toprule
        \textbf{Fig.} 
        & \textbf{AST} 
        & \textbf{Tokens}\\ \midrule
        \vspace{-0.3in}~$J$~\ref{fig:javaLoop} 
        & \includegraphics[width=2.8in]{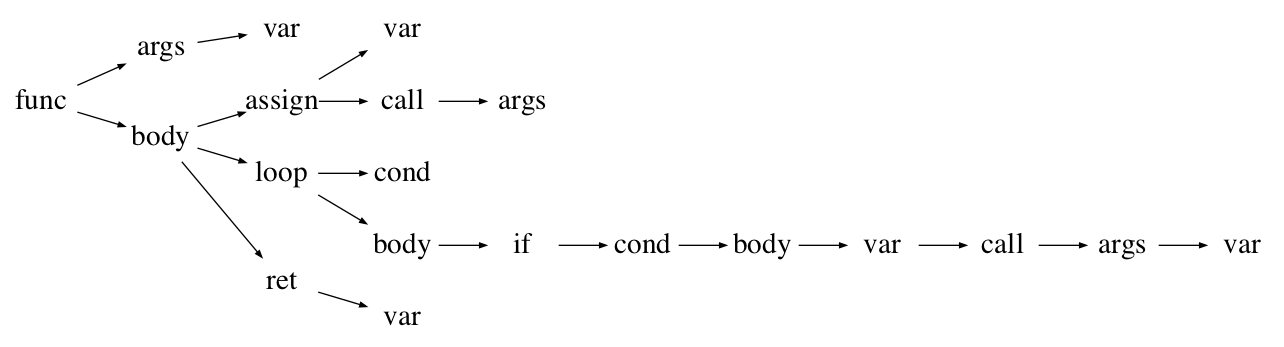}
        & \vspace{-0.3in}getevens, get, evens, max, list, integer, arraylist, array, add\\
        \vspace{-0.3in}$P$~\ref{fig:pyOdds} 
        & \includegraphics[width=2.8in]{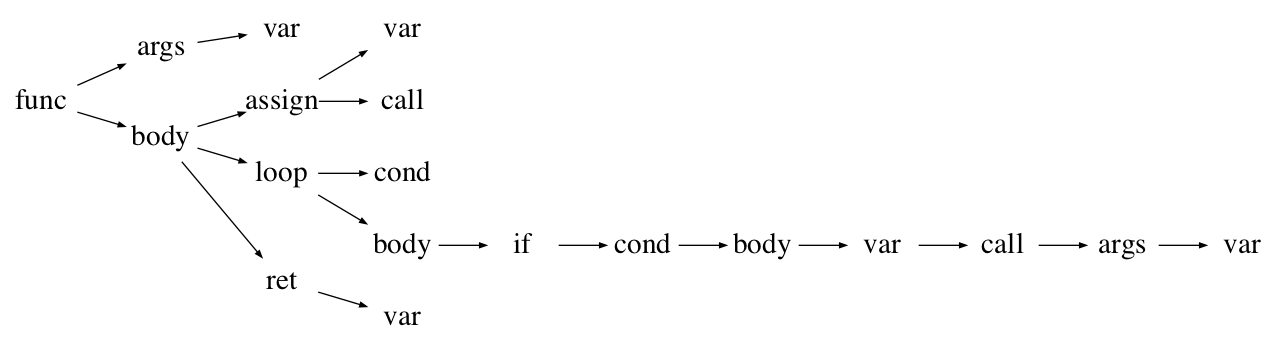}
        & \vspace{-0.4in}allodds, all, odds, range,  append\\
        \vspace{-0.3in}$J$~\ref{fig:javaStream} 
        & \includegraphics[width=3.1in]{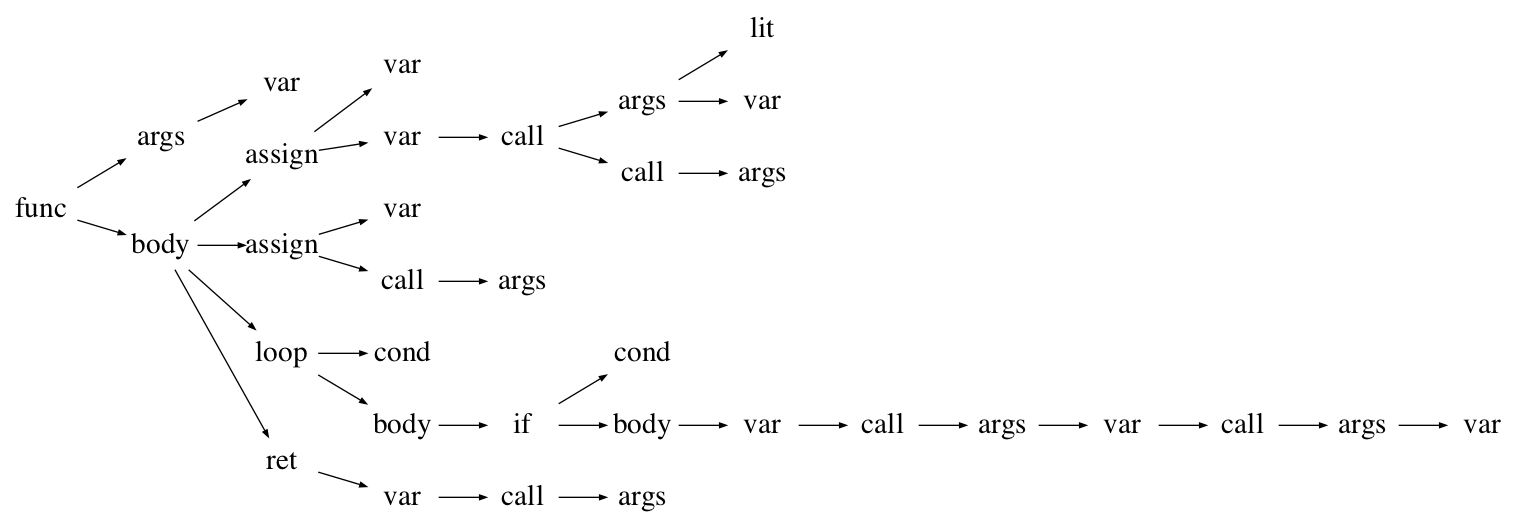}
        & \vspace{-0.1in}func, intstream, stream, range, toarray, array, list, integer, arraylist, length,get, add\\
        \vspace{-0.3in}$P$~\ref{fig:pyComp} 
        & \includegraphics[width=2.4in]{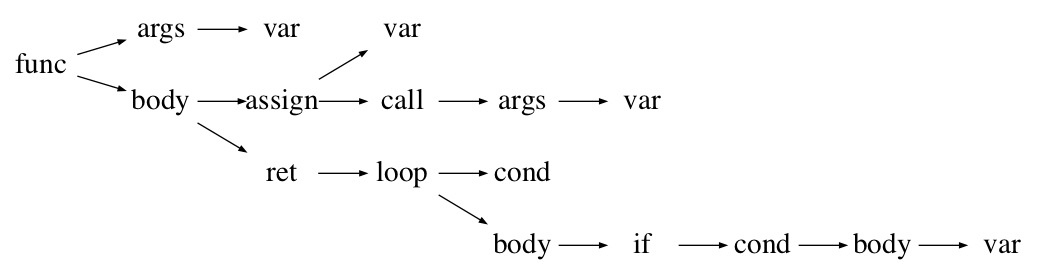}
        & \vspace{-0.2in}evennums, even, nums, maxval, max, val, xrange\\ \bottomrule
    \end{tabular}
         \caption{
     Generic ASTs and Tokens for \JAVA ($J$) and \PYTHON($P$) functions from \fref{codeSamples} \
    } \label{fig:tokensAndTrees}
\end{figure}

%% file: figs_mocs.tex
\begin{figure}
    \centering
    \includegraphics[width=\linewidth]{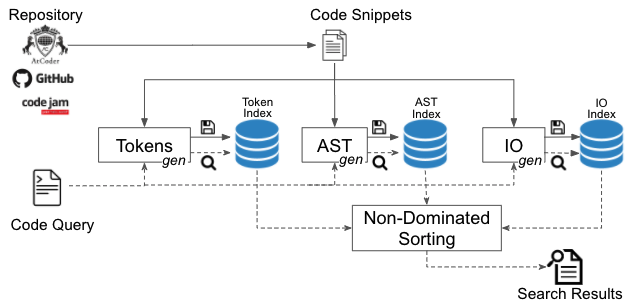}
    \caption{Overview of \approach{} }
    \label{fig:mocs}
\end{figure}

%% file: tab_metrics_comp.tex
\begin{table}[!t]
    \centering
    \small
    \caption{Similarity measures for \JAVA ($J$) and \PYTHON ($P$) functions from ~\fref{codeSamples}. High similarity implies high values ($\uparrow$) of \dToken{}, low values ($\downarrow$) of \dAST{}, and high values ($\uparrow$) of \dIO{}.}
    \label{tab:codeMetrics}
    \begin{tabular}{rrlrll}
        \toprule
        \textbf{Snip 1} & \textbf{Snip 2} & ${\bm{d_{token}}}$ ($\uparrow$) & $\bm{d_{AST}}$ ($\downarrow$) & \multicolumn{2}{c}{$\bm{d_{IO}}$ ($\uparrow$)} \\ \cmidrule(lr){5-6}
        &&&&\textbf{(s$_1$, s$_2$)}&\textbf{(s$_2$, s$_1$)}\\
        \midrule 
        $J$ - \ref{fig:javaLoop} & $P$ - \ref{fig:pyOdds} & 0.0 & 1 & 0.0 & 0.0 \\
        $J$ - \ref{fig:javaLoop} & $J$ - \ref{fig:javaStream} & 0.333 & 16 & 1.0 & 1.0\\
        $J$ - \ref{fig:javaLoop} & $P$ - \ref{fig:pyComp} & 0.067 & 7 & 1.0 & 1.0 \\
        $P$ - \ref{fig:pyOdds} & $J$ - \ref{fig:javaStream} & 0.0 & 17 & 0.5 & 0.3\\
        $P$ - \ref{fig:pyOdds} & $P$ - \ref{fig:pyComp} & 0.0 & 8 & 0.5 & 0.5\\
        $J$ - \ref{fig:javaStream} & $P$ - \ref{fig:pyComp} & 0.059 & 21 & 1.0 & 1.0\\
        \bottomrule
    \end{tabular}
\end{table}

%% file: tab_rqs.tex
\begin{table*}[tb]
    \centering
    \small
    \caption{
       Summary of RQs with the application (code-to-code search, clone detection), baselines, benchmarks (AtCoder or BigCloneBench) and the language(s). \approach{}$_<single>$ represents \approach{} using a single similarity measure.
    }
    \label{tab:rqDesc}
    \begin{tabular}{rp{2.3in}llll}
        \toprule
        \textbf{RQ} & \textbf{Purpose} & \textbf{Application} & \textbf{Baselines} & \textbf{Language(s)} & \textbf{Benchmarks} \\ \midrule
        1 & Merit of using multiple similarity measures & Code-to-Code Search & \approach$_{<single>}$ & \JAVA $\leftrightarrow$ \PYTHON & AtCoder \\
        2 & Vs state-of-the-practice \textbf{cross-language} tools & Code-to-Code Search & ElasticSearch, GitHub & \JAVA $\leftrightarrow$ \PYTHON & AtCoder \\
        3 & Vs state of the art \textbf{within-language} tools & Code-to-Code Search & \facoy{} & \JAVA & AtCoder, BigCloneBench \\
        4 & Using \approach{} to identify similar code & Clone Detection & \clcdsa{}, \astLearner{} & \JAVA $\leftrightarrow$ \PYTHON & AtCoder \\ \bottomrule
    \end{tabular}
\end{table*}

%% file: tab_data.tex
\begin{table}[tb]
    \centering
    \small
    \caption{Summaries of  AtCoder and BigCloneBench datasets}
    \vspace{-6pt}
    \begin{tabular}{lrrlr}
        \toprule
        \multicolumn{3}{c}{\textbf{AtCoder} (AtC)} & \multicolumn{2}{c}{\textbf{BigCloneBench} (BCB)} \\ 
        \textbf{Metric} & \textbf{\JAVA} & \textbf{\PYTHON} & \textbf{Metric} & \textbf{\JAVA}\\ \midrule
        \#Problems & 364 & 333 &  \#Features & 43\\
        \#Files & 20,828 & 22,318 & \#Files & 55,499 \\
        Avg. Files$/$Problem & 57 & 67 & Avg. Files$/$Feature & 1291 \\
        \#Methods & 81,896 & 10,020 & \#Methods & 765,331 \\
        Avg. Lines/File & 51 & 14 & Avg. Lines/File & 278 \\ \bottomrule
    \end{tabular}
    \label{tab:data}
\end{table}

%% file: tab_results_cross_language.tex
\begin{table}[tb]
    \centering
    \small
    \caption{
        RQ1 \& RQ2: Cross-language code search results on AtCoder dataset comparing \approach{} against the state-of-the-practice (\textit{SotP}) GitHub, and ElasticSearch. \approachToken{}, \approachAST{}, \approachSLACC{} use single search similarities (\textit{Single Sim.}) $d_{token}$, $d_{AST}$ and $d_{IO}$ respectively. \approachStatic{} uses $d_{token}$ and $d_{AST}$ with non-domination. \weightedApproach performs code search with KDTree using $d_{token}$, $d_{AST}$ and $d_{IO}$. Code search techniques using multiple  similarity measures are represented with \textit{Multi Sim.}
    }
    \label{tab:crossLangSearch}
    \begin{tabular}{m{1cm}p{1.8cm}ccc}
        \toprule
        &\textbf{Search} & \textbf{MRR} & \textbf{P@1/3/5/10} & \textbf{SR@1/3/5/10}\\ \midrule
        \multirow{2}{1cm}{\textit{SotP}} & ElasticSearch & 29 & 27/25/23/24 & 27/44/57/75 \\
        &GitHub & 37 & 32/36/38/39 & 32/49/60/73 \rowSep
        \multirow{3}{1cm}{\textit{Single Sim.}} & \approachToken{} & 31 & 27/31/40/42 & 27/48/58/72 \\
        & \approachAST{} & 34 & 34/41/45/44 & 34/41/58/82\\
        & \approachSLACC{} & 45 & 42/42/35/27 & 42/45/47/47\rowSep
        \multirow{3}{1cm}{\textit{Multi Sim.}}&\approachStatic{} & 43 & 40/45/44/48 & 40/72/85/86\\
        & \weightedApproach & 39 & 39/41/40/37 & 39/56/71/89\\
        & \approach{} & \textbf{64} & \textbf{58}/\textbf{64}/\textbf{65}/\textbf{61} & \textbf{58}/\textbf{88}/\textbf{91}/\textbf{94}\\
         \bottomrule
    \end{tabular}
\end{table}

%% file: tab_results_atc_bcb.tex
    

\begin{table}[tb]
    \centering
    \small
    \caption{
        RQ3: Single-language Java code search comparing \approach{} to the state-of-the-art (\textit{SotA}) \facoy{} on AtCoder and BigCloneBench.
    }
    \label{tab:sameLangSearch}
    \begin{tabular}{L{0.4cm}m{0.5cm}lccc}
        \toprule
        & & \textbf{Search} & \textbf{MRR} & \textbf{P@1/3/5/10} & \textbf{SR@1/3/5/10}\\ \midrule 
        \multirow{6}{0.2cm}{\rotatebox[origin=c]{90}{\textbf{AtCoder}}} & \textit{SotA} & \facoy{} & 51 & 37/35/33/32 & 37/40/49/63 \cRowSep{2-6}
        & \multirow{3}{1cm}{\textit{Single Sim.}} & \approach{}$_{tokens}$ & 46 & 36/32/31/29 & 36/40/45/58\\
        & &\approachAST{} & 40 & 38/33/31/28 & 38/42/51/69\\
        & &\approachSLACC{} & 40 & 39/39/38/32 & 39/48/52/59 \cRowSep{2-6}
        &\vspace{-0.3cm}\multirow{3}{1cm}{\textit{Multi Sim.}} & \approach{}$_{static}$ & 53 & 43/45/44/41 & 43/58/65/77\\
        & & \approach{} & \textbf{57} & \textbf{50}/\textbf{53}/\textbf{54}/\textbf{48} &
        \textbf{50}/\textbf{63}/\textbf{75}/\textbf{88}\\\midrule
        \multirow{6}{0.2cm}{\rotatebox[origin=c]{90}{\textbf{BigCloneBench}}} & \textit{SotA} & FaCoY & 76 & 70/68/68/65 & 70/72/74/81 \cRowSep{2-6}
        & \multirow{3}{1cm}{\textit{Single Sim.}} & \approach{}$_{tokens}$ & 75 & 69/65/61/59 & 69/72/74/81 \\
        & &\approach{}$_{AST}$ & 72 & 68/61/55/51 & 68/74/76/83 \\
        & &\approach{}$_{\slacc{}}$ & 07 & 06/02/01/01 & 06/07/07/09 \cRowSep{2-6}
        & \multirow{2}{1cm}{\textit{Multi Sim.}} & \approach{}$_{static}$ & \textbf{81} & 76/\textbf{73}/\textbf{72}/67 & 76/\textbf{81}/\textbf{89}/\textbf{94} \\
        & &\approach{} & \textbf{81} & \textbf{77}/\textbf{73}/\textbf{72}/\textbf{68} & \textbf{77}/\textbf{81}/\textbf{89}/\textbf{94} \\\bottomrule
    \end{tabular}
    ~\vspace{-12pt}
\end{table}

%% file: tab_results_clones.tex
\begin{table}[tb]
    \centering
    \small
    \caption{
        RQ4: Cross-language performance of \approach{} in clone detection compared to \astLearner{}, CLCDSA, and SLACC on AtCoder.
    }
    \label{tab:crossLangClone}
    \begin{center}
    \begin{tabular}{m{1.5cm}lccc}
        \toprule
        & \textbf{Clone Detector} & \textbf{Precision} & \textbf{Recall} & \textbf{F1}\\ \midrule
        \multirow{3}{1.5cm}{\textit{Single Sim.}} & ASTLearner & 25 & 80 & 38\\
        & CLCDSA & 49 & 83 & 62 \\
         & \slacc{} & \textbf{66} & 19 & 30 \rowSep{}
        \multirow{2}{1.5cm}{\textit{Multi Sim.}} & \approach{}$_{static}$ & 48 & 85 & 61\\
        & \approach{} & 55 & \textbf{89} & \textbf{68}\\ \bottomrule
    \end{tabular}
    \end{center}
\end{table}

%% file: tab_semantics_discussion.tex
\begin{table}[tb]
    \centering
    \small
    \caption{
        Performance based on 4,984 executable code snippets from BigCloneBench.
    }
    \label{tab:bcbSemantic}
    \begin{tabular}{m{1cm}p{1.6cm}ccc}
        \toprule
        &\textbf{Search} & \textbf{MRR} & \textbf{P@1/3/5/10} & \textbf{SR@1/3/5/10}\\ \midrule
        \textit{SotP} & GitHub & 68 & 64/58/54/46 & 64/68/72/75 \rowSep{}
        \textit{SotA} & FaCoY & 79 & 74/70/68/57 & 74/76/81/84 \rowSep{}
       \textit{Single Sim.} & \approach{}$_{\slacc{}}$ & 82 & \textbf{81}/78/74/67 & \textbf{81}/83/89/94 \rowSep{}
        \multirow{2}{1cm}{\textit{Multi Sim.}} & \approach{}$_{static}$ & 80 & 78/75/72/66 & 79/83/87/91 \\
        & \approach{} & \textbf{83} & \textbf{81}/\textbf{79}/\textbf{74}/\textbf{68} & \textbf{81}/\textbf{86}/\textbf{91}/\textbf{96} \\\bottomrule
    \end{tabular}
\end{table}

%% file: tab_correlation.tex
\begin{table}[tb]
    \centering
    \small
    \caption{\added{Pearson's correlation ($r$) between \dToken{}, \dAST{} and \dIO{} for cross-language snippets on AtCoder (AtC) and within-language Java snippets on AtCoder and on 4,984 executable BigCloneBench(BCB) datasets.}}
    \begin{tabular}{ccccc}
    \toprule
    \multirow{2}{*}{\textbf{Dataset}} & \multirow{2}{*}{\textbf{Language}}  & \multicolumn{3}{c}{\textbf{Correlations} ($r$)} \cRowSep{3-5}
    & & $token,AST$ & $token,IO$ & $AST,IO$ \\ \midrule
    AtC & \JAVA{} $\leftrightarrow$ \PYTHON{} & -0.38 & 0.33 & -0.41 \\
    AtC & \JAVA{} $\leftrightarrow$ \JAVA{} & -0.49 & 0.51 & -0.68 \\
    BCB & \JAVA{} $\leftrightarrow$ \JAVA{} & -0.46 & 0.53 & -0.71 \\ \bottomrule
    \end{tabular}
    \label{tab:correlation}
\end{table}

%% file: figs_os_code_samples.tex
\captionsetup[subfigure]{labelfont=bf,textfont={bf},singlelinecheck=off,margin=12pt,skip=2pt}

\begin{figure}[!t]
    \begin{subfigure}{\textwidth}
    \begin{minted}[xleftmargin=2em,linenos,firstnumber=1,fontsize=\footnotesize]{Java}
class HashMultiSet<E> ... {
  ...
  public int count(Object element) {
    Count frequency = Maps.safeGet(backingMap, element);
    return (frequency == null) ? 0 : frequency.get();
  }
  ...
}
    \end{minted}
    \caption{Method that returns count of a MultiSet from \mintJ{google-guava}}
    \label{fig:guavaCountMultiSet}
    \end{subfigure}
    \medskip
    \begin{subfigure}{\textwidth}
    \begin{minted}[xleftmargin=2em,linenos,firstnumber=1,fontsize=\footnotesize]{Python}
class Counter(dict):
  """
  ... count ...
  """
  ...
  def __getitem__(key):
    return self.get(key, 0)
  ...
    \end{minted}
    \caption{Function to get count of a key from a \mintP{Counter} from\\ \mintP{collections} library.}
    \vspace{-12pt}
    \label{fig:collectionsCount}
    \end{subfigure}
    \caption{Open Source code; the query in (a) yields (b)
    based on cross-language static and dynamic information}
    \label{fig:osCodeSamples}
    \vspace{-12pt}
\end{figure}